\documentclass[a4paper,prd,showpacs,noshowkeys,preprint]{revtex4}
\usepackage[latin1]{inputenc}
\usepackage{amsmath}
\usepackage{amssymb}
\usepackage{mathrsfs}

\providecommand{\bp}{\mathbf{p}}
\providecommand{\rp}{\mathrm{p}}
\providecommand{\bk}{\mathbf{k}}
\providecommand{\rk}{\mathrm{k}}
\providecommand{\bq}{\mathbf{q}}
\providecommand{\q}{\mathrm{q}}
\providecommand{\bx}{\mathbf{x}}
\providecommand{\rx}{\mathrm{x}}
\providecommand{\lag}{\mathscr{L}}

\providecommand{\sla}[1]{~\hs{-1.5ex}\not\hs{-.4ex}#1\hs{.1ex}}

\providecommand{\sww}{_{_{\rm WW}}}
\providecommand{\pe}[1]{#1{\cdot}}

\providecommand{\eq}[1]{\begin{equation} #1 \end{equation}}
\providecommand{\eqarr}[1]{\begin{eqnarray} #1 \end{eqnarray}}
\providecommand{\ket}[1]{\vert #1 \rangle}
\providecommand{\bra}[1]{\langle #1 \vert}
\providecommand{\braket}[2]{\langle #1\vert #2 \rangle}
\providecommand{\aver}[1]{\langle #1 \rangle}
\providecommand{\ms}[1]{\mbox{\small $#1$}}
\providecommand{\hs}[1]{\hspace{#1}}
\providecommand{\bs}[1]{\boldsymbol{#1}}
\font\bb=bbmss12 scaled 1000
\def\id{\mbox{\bb 1}}
\begin{document}
%%%%%%%%%%%%%%%%%%%%%%%%%%%%%%%%%%%%%%%%%%%%%%%%%
\title{Intrinsic flavor violation for massive neutrinos}
% \author{C.~C.~Nishi\,$^{\rm(ab)}$}
\author{C.~C.~Nishi}
\email{ccnishi@ifi.unicamp.br}
% \author{M.~M.~Guzzo\,$^{\rm(a)}$}
% \email{guzzo@ifi.unicamp.br}
\affiliation{
%$^{\rm(a)}$
Instituto de Física ``Gleb Wataghin''\\
Universidade Estadual de Campinas, Unicamp\\
13083-970, Campinas, SP, Brasil
}
\affiliation{
%$^{\rm(b)}$
Instituto de Física Teórica,
UNESP -- São Paulo State University\\
Rua Pamplona, 145,
01405-900 -- São Paulo, Brasil
}

%\date{\today}
%%%%%%%%%%%%%%%%%%%%%%%%%%%%%%%%%%%%%%%%%%%%%%%%%
\begin{abstract}
It is shown that intrinsic neutrino flavor violation invariably occurs when
neutrinos are created within the SM augmented by the known massive neutrinos, with
mixing and nondegenerate masses.
The effects are very small but much greater than the naive estimate $\Delta
m^2/E_\nu^2$ or the branching ratio of indirect flavor violating processes such as
$\mu\rightarrow e\gamma$ within the SM. We specifically calculate the probability
(branching ratio) of pion decay processes with flavor violation, such as
$\pi\rightarrow \mu\bar{\nu}_e$, showing nonzero results.

\end{abstract}
%%%%%%%%%%%%%%%%%%%%%%%%%%%%%%%%%%%%%%%%%%%%%%%%%
%\keywords{ }
%\twocolumn
% 14.60.Pq Neutrino mass and mixing
% 13.15.+g Neutrino interactions
% 12.15.F Quark and lepton masses and mixing
% 11.10.-z Field theory
\maketitle
%%%%%%%%%%%%%%%%%%%%%%%%%%%%%%%%%%%%%%%%%%%%%%%%%
\section{Introduction}
\label{sec:intro}

After the confirmation that neutrinos are massive, nondegenerate and mix themselves,
further investigations are being intensively carried out, experimentally as well as
theoretically, to clarify the remaining mysteries about the neutrinos and the new
physics they could be hiding\,\cite{smirnov:window}. One question that massive
neutrinos immediately poses concerns the status of lepton number ($L$) and family
lepton numbers ($L_e,L_\mu,L_\tau$) that were automatically conserved in the
standard model (SM) without right-handed singlet neutrinos.
We know from the successful observation of neutrino flavor oscillations that family
lepton numbers are not conserved quantities due to the presence of the
nondiagonal MNS mixing matrix. Total lepton number could be conserved at the
classical level if neutrinos were Dirac fermions but that scenario does not explain
the smallness of neutrino masses. The most natural way to explain tiny neutrino
masses is the seesaw mechanism but, in this case, neutrinos are Majorana fermions in
general. Although, approximate lepton number conservation can be achieved,
guaranteeing small active neutrino masses, by assigning appropriate lepton numbers to
heavy SM gauge singlets\,\cite{kersten}. 
% To confirm the Majorana nature of neutrinos, the observation of neutrinoless
% double beta decay is crucial. 

Indeed, it is exactly in the seesaw scenario that many interesting physics could be
potentially observable. If the seesaw scale is relatively low, at the order of TeV,
effects such as the violation of unitarity of the MNS matrix\,\cite{xing:unitvio} may
be observable or the direct production of heavy seesaw
particles\,\cite{kersten,xing:typeII}, including heavy neutrinos (type I or III) or
heavy scalars (type II), might be possible. Non-standard interactions could also
modify the standard oscillation formulas\,\cite{kopp}. In such context, it is common
to think that all consequences of the SM augmented by massive neutrinos have been
investigated through. (An extensive analysis can be found in
Ref.\,\onlinecite{shrock}.) Most of the direct consequences of massive neutrinos,
with the exception of neutrino oscillations, are very difficult to be observed due to
the tiny masses and mass differences: $|\Delta m^2_{12}|\approx 8\times 10^{-5}\rm
eV^2$ and $|\Delta m^2_{23}|\approx 2.3\times 10^{-3}\rm
eV^2$\,\cite{vissani:review}. For example, the production of antineutrinos with
negative helicity is possible in principle, because neutrinos are massive, but
negligible in practice\,\cite{pi:helicity}. Despite such difficulties, an enormous
experimental effort is being dispended to measure the absolute neutrino mass
scale\,\cite{giunti}. On the other hand, indirect effects allowed by massive
neutrinos with mixing, such as the lepton flavor (LF) violating decay $\mu\rightarrow
e\gamma$, are even strongly suppressed in the SM [Br($\mu\rightarrow
e\gamma)<10^{-50}$] because of the tiny neutrino masses that enter the
loops\,\cite{bilenky,casasibarra}. Extensions of the SM, though, may lead in general
to relatively large LF violating effects and certain conditions should be fulfilled
for a natural suppression\,\cite{LFV}.

Contrary to usual expectations, we will show in this article that intrinsic neutrino
flavor violation, hence lepton flavor violation, is possible in neutrino creation due
solely to the known neutrino mass differences and nonzero mixing. More specifically,
we will show that processes such as $\pi\rightarrow\mu\bar{\nu}_e$, are possible with
a branching ratio much greater than loop induced processes such as $\mu\rightarrow
e\gamma$. In fact, this effect should be correctly quantified before considering
new physics contributions that could mimic the same
effects\,\cite{grossman:NI,johnson:NI}. For instance, there were attempts to
explain the LSND anomaly\,\cite{lsnd} from new physics interactions that violate
lepton flavor\,\cite{lsnd:Lvio}. For interactions that conserve total lepton
number, however, conflicts with low energy phenomena can not be
avoided\,\cite{grossman:lsnd}. Before the confirmation that neutrino oscillations
were responsible for both deficits of neutrinos coming from the sun and the
atmosphere, there were attempts to explain the deficit with non-standard
interactions\,\cite{grossman:atm}, even with massless neutrinos\cite{guzzo}.
Indeed, it is important to distinguish the intrinsic lepton flavor violation effect
calculated here from effects coming from interactions, extrinsic to the presence of
neutrino masses, that violate lepton flavor and, perhaps, lepton number. Such
interactions could give rise to effective operators with observable consequences in
other low energy phenomena. An analogous distinction between direct and indirect CP
violation is important to classify the CP violating effects involving the neutral
$K$-mesons\,\cite{indirectCP} that confirmed the CKM mechanism of CP violation in the
SM\,\cite{km}.

The outline of the article is as follows: in Sec.\,\ref{sec:WW} we apply the
Wigner-Weisskopf approximation to treat the pion decay, considering the finite decay
width. Section\,\ref{sec:flavorvio} contains the main results of neutrino flavor
violation in pion decay and uses mainly Eqs.\,\eqref{dchidt1} and \eqref{M+dM} from
Sec.\,\ref{sec:WW}. The ones only interested in the results may skip
Sec.\,\ref{sec:WW}. We discuss the results and some implications in
Sec.\,\ref{sec:discussion}. The appendices show some calculations that were omitted
through the text and some useful material.

%%%%%%%%%%%%%%%%%%%%%%%%%%%%%%%%%%%%%%%%%%%
\section{Wigner-Weisskopf approximation in pion decay}
\label{sec:WW}

Consider the pion decay $\pi^-\rightarrow l_i^-+\bar{\nu}_j$, $i=1,2$ ($l_1\equiv
e,l_2\equiv \mu$) and $j=1,2,3$.
The detailed description of this decay will be made by applying the Wigner-Weisskopf
(WW) approximation method\,\cite{WW}. The WW method is essentially an improved method
of second order time dependent perturbation theory which can describe the dynamics of
decaying and decayed states at intermediate times (exponential behavior).

To calculate the decaying pion state at any time $t$, within the applicable
approximation that only $l_i\bar{\nu}_j$ states appear as decay states, it
suffices to discover the functions $\psi$ and $\chi$ in
\eq{
\label{pi(t)}
\ket{\pi(t)}\sww=
\int\! d^3\rp\,\psi(\bp,t)e^{-iE_\pi t}\ket{\pi(\bp)}
+\sum_{ij}\int\! d^3\q d^3\rk\,\chi_{ij}(\bq,\bk;t)\,e^{-i(E_{l_i}+E_{\nu_j})t}
\ket{l_i(\bq)\nu_j(\bk)}~,
}
where the spin degrees of freedom are omitted and the states
$\{\ket{\pi(\bp)},\ket{l_i(\bq)\nu_j(\bk)}\}$, $i,j=1,2$, refer to the free
states, eigenstates of $H_0$, normalized as
\eqarr{
\label{norm}
\braket{\pi(\bp')}{\pi(\bp)}&=&\delta^3(\bp-\bp')~,\cr
\braket{l_i(\bq')\nu_j(\bk')}{l_i(\bq)\nu_j(\bk)}
&=&\delta^3(\bq-\bq')\delta^3(\bk-\bk')
~.
}
The expansion \eqref{pi(t)} means we are restricted to
the lowest order of perturbation theory.

The free Hamiltonian is characterized by the free energy of the states with
physical masses
\eqarr{
H_0\ket{\pi(\bp)}&=&E_\pi(\bp)\ket{\pi(\bp)}~,
\\
H_0\ket{l_i(\bq)\nu_j(\bk)}&=&\big(E_{l_i}(\bq)+E_{\nu_j}(\bk)\big)
\ket{l_i(\bq)\nu_j(\bk)}
~,
}
where $E_\alpha(\bp)=\sqrt{\bp^2+M_\alpha^2}$ ($\alpha=\pi,l_i,\nu_j$), 
and we will denote $M_{l_i}\equiv M_i$ and $M_{\nu_j}\equiv m_j$.
The interaction Hamiltonian is given by
\eq{
V= -\int d^3\rx \lag_F(\bx) + \text{counter terms}~,
}
where $\lag_F$ is the Fermi interaction Lagrangian.

Considering the total Hamiltonian 
\eq{
H=H_0+V\,,
}
we can write a Schrödinger-like equation
\eqarr{
(i\frac{d}{dt}-H_0)\ket{\pi(t)}\sww&=&
\int d^3\rp\,i\frac{\partial \psi(\bp,t)}{\partial t}e^{-iE_\pi t}\ket{\pi(\bp)}
\cr
&&
+\sum_{ij}\int\! d^3\q d^3\rk\,
i\frac{\partial \chi_{ij}(\bq,\bk;t)}{\partial t}\,e^{-i(E_{l_i}+E_{\nu_j})t}
\ket{l_i(\bq)\nu_j(\bk)}~,
\\&=&
V\ket{\Psi(t)}~.
}
Contraction with the appropriate states yields
\eqarr{
\label{dpsidt1}
i\frac{\partial}{\partial t}\psi(\bp,t)&=&\frac{\delta M^2}{2E_\pi}
\psi(\bp,t)+ \sum_{ij}\int\! d^3\q d^3\rk\,
\chi_{ij}(\bq,\bk;t)\,
\bra{\pi(\bp)}V(t)\ket{l_i(\bq)\nu_j(\bk)}~,
\\
\label{dchidt1}
i\frac{\partial}{\partial t}\chi_{ij}(\bq,\bk;t)&=&
\int d^3\rp \psi(\bp,t)
\bra{l_i(\bq)\nu_j(\bk)}V(t)\ket{\pi(\bp)}~,
}
where $V(t)=e^{iH_0t}Ve^{-iH_0t}$ and $\delta M^2$ is a counter term.

%  ~~~[\int d^3\rp |\psi(\bp)|^2=1]\\

From the initial conditions
\eqarr{
\label{ic:psi}
\psi(\bp,0)&=&\psi(\bp)\,,
\\
\label{ic:chi}
\chi_{ij}(\bq,\bk;0)&=&0~,
}
we can formally solve
\eq{
\chi_{ij}(\bq,\bk;t)=
-i\int_{0}^{t}\!dt'
\int \!d^3\rp\, \psi(\bp,t')
\bra{l_i(\bq)\nu_j(\bk)}V(t')\ket{\pi(\bp)}~,
}
and obtain
\eqarr{
\label{dpsidt2}
\frac{\partial}{\partial t}\psi(\bp,t)&=& -i\frac{\delta M^2}{2E_\pi}
\psi(\bp,t)
+ \int\!d^3\rp' d^3\q d^3\rk\int_{0}^{t}\!dt'\,
\bra{\pi(\bp)}V(t)\ket{l_i(\bq)\nu_j(\bk)}
\cr &&
\phantom{-i\frac{\delta M^2}{2E_\pi}\psi(\bp,t) +\int\!d^3\rp'}
\times
\bra{l_i(\bq)\nu_j(\bk)}V(t')\ket{\pi(\bp')}\psi(\bp',t')
~.
}
This is the key equation for the WW approximation. 
% Note that from the initial condition \eqref{ic:chi} we obtain
% $\frac{\partial}{\partial t}(\psi(\bp,t)e^{-i\delta Et})|_{t=0}=0$ which gives the
% short time $1-(~)t^2$ behavior of $|\psi(\bp,t)|^2$.

Notice that only momentum conservation holds for the matrix elements, in particular,
\eq{
\label{<V>}
\bra{l_i(\bq)\nu_j(\bk)}V\ket{\pi(\bp)}=
N_{ij}^{-1/2}
\mathscr{M}_{ij}
\,\delta^3(\bp-\bq-\bk)~,
}
where
$N_{ij}=(2\pi)^3 2E_{l_i}(\bq)2E_{\nu_j}(\bk)2E_\pi(\bp)$ and
$\mathscr{M}_{ij}\equiv\mathscr{M}_{ij}(\bp,\bq,\bk)=
\mathscr{M}(\ms{\pi^-(\bp)\rightarrow l_i^-(\bq)\bar{\nu}_j(\bk)})
$.
Replacing Eq.\,\eqref{<V>} into Eq.\,\eqref{dpsidt2} yields
\eq{
\label{dpsidt3}
\frac{\partial}{\partial t}\psi(\bp,t)=
-i\frac{\delta M^2}{2E_\pi} \psi(\bp,t)
-\frac{1}{2E_\pi(\bp)}\int_0^{t}\!dt'\,
\psi(\bp,t-t')K(\bp,t')
~,
}
where
\eq{
\label{K}
K(\bp,t')=\frac{1}{(2\pi)^3}\sum_{ij}
\int \frac{d^3\q}{2E_{l_i}}\frac{d^3\rk}{2E_{\nu_j}}
e^{i\Delta E_{ij}t'}
|\mathscr{M}_{ij}|^2
\delta^3(\bp-\bq-\bk)~,
}
where $\Delta E_{ij}\equiv E_\pi-E_{l_i}-E_{\nu_j}$ and the respective $\bp,\bq,\bk$
dependence of $E_\pi,E_{l_i},E_{\nu_j}$ is implicit.
The expression in Eq.\,\eqref{K}, however, does not provide a convergent
integral since $|\mathscr{M}_{ij}|^2$ behaves as $\bk^2$ for 
$\bq=\bp-\bk$ and $|\bk|\rightarrow\infty$.
However, a cutoff function $f(\bp,\bq,\bk)$ multiplying $\mathscr{M}_{ij}$ is
understood to regularize the expression. Such function can arise effectively from the
pion form factor and vertex corrections in higher orders\,\cite{DGH}. Such cutoff
function is necessary to ensure the convergence of Eq.\,\eqref{K} and the production
rate of $\pi(\bp)\rightarrow l_i(\bq)\bar{\nu}_j(\bk)$ to be more probable for the
energy conserving states and do not grow indefinitely for high $|\bk|$.
We will assume that the cutoff function $f$ satisfies the properties
\begin{itemize}
\item[(P1)] the functional form of $f$ is broad for
$E_{l_i}$ or $E_{\nu_j}$ and it varies very slowly for values close to the energy
conserving values, in particular $f=1$ for $\Delta E_{ij}=0$.
\item[(P2)] the suppression of high momentum $|\bk|$ or $|\bq|$ (with $\bq+\bk$
fixed) occurs only significantly at an scale $\Lambda$ which satisfies
$\Gamma\ll\Lambda\ll M^2_\pi/\Gamma$, where $\Gamma$ is the pion decay width.
\end{itemize}
Only these properties will be necessary for most of the calculations in this article.
The inclusion of an explicit cutoff function will be considered in appendix
\ref{ap:cutoff} to justify the property (P2).

With the introduction of $f$ we can argue that the dominant contribution of
$K(\bp,t)$ is for $t \sim 0$, since Eq.\,\eqref{K} corresponds to a Fourier transform
in $E_{\nu_j}$ and the integrand is a very broad function, which leads to a narrow
function in time. We can then approximate Eq.\,\eqref{dpsidt2} as
\eqarr{
\label{dpsidt4}
\frac{\partial}{\partial t}\psi(\bp,t)\approx
-i\frac{\delta M^2}{2E_\pi} \psi(\bp,t)
-\frac{1}{2E_\pi(\bp)}\left[\int_0^{\infty}\!dt'\,K(\bp,t')\right]
\psi(\bp,t)
~.
}
The Eq.\,\eqref{dpsidt4} corresponds to the WW approximation and it is valid
for intermediate times, i.e., $t$ should be greater than the time width of
$K(\bp,t)$, since for such short time the original expression \eqref{dpsidt3}
can be significantly different. 
% The upper bound ...
Within the WW approximation the expression inside the bracket in Eq.\,\eqref{dpsidt4}
gives
\eq{
\label{intK}
\int_0^{\infty}\!dt'\,K(\bp,t')
=
\frac{i}{(2\pi)^3}\sum_{ij}
\int \frac{d^3\q}{2E_{l_i}}\frac{d^3\rk}{2E_{\nu_j}}
\frac{|f\mathscr{M}_{ij}|^2}
{\Delta E_{ij}+i\epsilon}
\delta^3(\bp-\bq-\bk)~.
}
Using the relation
\eq{
\frac{1}{E\pm i\epsilon}=\mathcal{P}\frac{1}{E}\mp i\pi\delta(E)~,
}
we obtain
\eqarr{
\label{ReK}
\mathrm{Re}\,\text{Eq.\,\eqref{intK}}&=&
\frac{\pi}{(2\pi)^3}\sum_{ij}
\int \frac{d^3\q}{2E_{l_i}}\frac{d^3\rk}{2E_{\nu_j}}
|f\mathscr{M}_{ij}|^2
\delta^4(\bp-\bq-\bk)~,
\\
\label{ImK}
\mathrm{Im}\,\text{Eq.\,\eqref{intK}}&=&
\frac{1}{(2\pi)^3}\sum_{ij}
\mathcal{P}\int \frac{d^3\q}{2E_{l_i}}\frac{d^3\rk}{2E_{\nu_j}}
\frac{|f\mathscr{M}_{ij}|^2}{\Delta E_{ij}}
\delta^3(\bp-\bq-\bk)~.
}
Using the property (P1) of $f$ we can identify Eq.\,\eqref{ReK} as proportional
to the pion decay rate at rest\,\cite{DGH}
\eq{
\mathrm{Re}\,\text{Eq.\,\eqref{intK}}=M_\pi \Gamma
~,
}
while Eq.\,\eqref{ImK} can be absorbed by the counterterm
\eq{
\mathrm{Re}\,\text{Eq.\,\eqref{intK}}= -\delta M^2~.
}

We can finally find the functions $\psi$ and $\chi$.
Equation \eqref{dpsidt4} gives
\eq{
\frac{\partial}{\partial t}\psi(\bp,t)=
-\frac{\Gamma}{2\gamma}\psi(\bp,t)~,
}
which can be readily solved to give
\eq{
\psi(\bp,t)=\psi(\bp)e^{-\Gamma t/2\gamma}~,
}
in accordance to the expected exponential decay law.
The factor $\gamma=E_\pi(\bp)/M_\pi$ accounts for the Lorentz dilatation of
time.
At the same time, the production wave function can be obtained from
Eq.\,\eqref{dchidt1}
\eqarr{
\chi_{ij}(\bq,\bk;t)&=&
\tilde{\chi}_{ij}(\bp,\bq,\bk;t)
\,\psi(\bp)\big|_{\bp=\bq+\bk}
\,,
\\
\tilde{\chi}_{ij}(\bp,\bq,\bk;t)
&\equiv&
\big[1-e^{-i(\Delta E_{ij}-i\Gamma/2\gamma)t}\big]
N_{ij}^{-1/2}
\frac{f\mathscr{M}_{ij}(\bp,\bq,\bk)}
{\Delta E_{ij}-i\frac{\Gamma}{2\gamma}}
\,.
}
Thus $|\chi_{ij}(\bq,\bk;t)|^2$ is the production probability density.

From the conservation of probability at any time $t$, we must check if
\eq{
\label{psi+chi}
\int \!d^3\rp|\psi(\bp,t)|^2
+\sum_{ij}\int \!d^3\q d^3\rk|\chi_{ij}(\bq,\bk;t)|^2
=1\,.
}
The calculation is performed in appendix \ref{ap:pi+lnu=1}. The important point is
that Eq.\,\eqref{psi+chi} is satisfied if we neglect the terms that does not
conserve energy in the squared amplitude $|\mathscr{M}_{ij}|^2$, i.e., the second
term in 
\eq{
\label{M+dM}
\sum_{\rm spins}|\mathscr{M}_{ij}|^2=|\mathscr{M}_{ij}^{\rm EC}|^2+
|\delta\mathscr{M}_{ij}|^2
\,,
}
where the upperscript EC stands for energy conservation. 
Notice that the usual energy conserving term $|\mathscr{M}_{ij}^{\rm EC}|^2$ is
positive definite while $|\delta\mathscr{M}_{ij}|^2$ has no definite sign. The
cutoff function $f$ is responsible for controlling such contributions. Therefore
we retain only the energy conserving parts of $|\mathscr{M}_{ij}|^2$ further on.

For future use, we also define 
\eq{
\label{Gammaij}
M_\pi\Gamma_{ij}=
\frac{\pi}{(2\pi)^3}|\mathscr{M}_{ij}^{\rm EC}|^2
\int\! d\Omega_k\Big[
\Big(\frac{k^2}{2E_{l_i}2E_{\nu_j}}
\Big(\frac{d(E_{l_i}+E_{\nu_j})}{dk}\Big)^{-1}
\Big]_{\rm EC}
\,,
}
and
\eq{
\label{Gammai}
\Gamma_{i}=\sum_j\Gamma_{ij}\,.
}
The ratio $\Gamma_i/\Gamma$ corresponds to the branching ratio of the reaction
$\pi\rightarrow l_i+\bar{\nu}$, independent of neutrino flavor, and it practically
coincides with the usual branching ratio calculated with massless neutrinos, since
$\sum_j|U_{ij}|^2=1$ and the kinematical contribution of neutrino masses are
negligible. Obviously, $\sum_i\Gamma_i=\Gamma$.

As a last remark, we should emphasize that nowhere in this section the precise form
of the interaction was used, except in the asymptotic behavior of
$|\mathscr{M}_{ij}|^2$. Therefore, this approximation can be used in any two-body
decay for which the interaction Hamiltonian is known, as long as a proper cutoff
function is understood. The explicit amplitude $\mathscr{M}_{ij}$ and squared
amplitude $|\mathscr{M}_{ij}|^2$ for pion decay are shown in appendix
\ref{ap:pidecay}.

%%%%%%%%%%%%%%%%%%%%%%%%%%%%%%%%%%%%%%%%%%%
\section{Neutrino flavor violation in pion decay}
\label{sec:flavorvio}

We begin by defining the two-particle states with definite flavor\,\cite{endnote1}
\eq{
\label{alphabeta}
\ket{l_\alpha(\bq)\bar{\nu}_\beta(\bk)}\equiv
\delta_{\alpha i}U_{\beta j} \ket{l_i(\bq)\bar{\nu}_j(\bk)}
\,.
}
The charged lepton states remain as mass eigenstates while the neutrino states are
mixed through $U_{\beta j}$. We will see, in accordance to usual expectations, that
pions decay mainly into the states $\ket{l_\alpha(\bq)\bar{\nu}_\beta(\bk)}$ with
$(\alpha,\beta)=(\mu,\mu)$. However, we will also see that there is a
non-null probability of the pion to decay into the neutrino flavor violating states
with $(\alpha,\beta)=(\mu,e)$ or $(\alpha,\beta)=(e,\mu)$.
For that purpose, we want to ultimately calculate the probability
\eq{
\label{Plnu:0}
\mathcal{P}_{l_\alpha\nu_\beta}(t)=
\int\! d^3\!\bq\int\! d^3\bk\, \sum_{\rm spins}
|\braket{l_\alpha(\bq)\bar{\nu}_\beta(\bk)}{\pi(t)}\sww|^2
\,.
}

Using $\chi_{ij}$ in Eq.\,\eqref{dchidt1},  when $t\gg 1/\Gamma$, we can rewrite
Eq.\,\eqref{Plnu:0} as
\eq{
\label{Plnu}
\mathcal{P}_{l_\alpha\nu_\beta}(t)=
\int\! d^3\!\bp\,|\psi(\bp)|^2\!
\int\! d^3\bk\, \sum_{\rm spins}
\Big|\sum_j
U_{\alpha j}e^{-iE_{\nu_j}t}U^{\dag}_{j\beta}F_{\alpha j}
\Big|^2_{\bq=\bp-\bk}
\,,
}
where
\eq{
\label{F:def}
U_{\alpha j}F_{\alpha j}(\bp,\bq,\bk)\equiv
N_{\alpha j}^{-1/2}\frac{f\mathscr{M}_{\alpha j}(\bp,\bq,\bk)}
{\displaystyle\Delta E_{\alpha j}-i\frac{\Gamma}{2\gamma}}
\,.
}
We see the exponential $e^{-iE_{\nu_j}t}$ is responsible for the neutrino
oscillation phenomenon. In fact, if we neglect the neutrino mass $m_j$ in every term
of Eq.\,\eqref{Plnu}, except in the exponential, we get
\eq{
\label{Plnu:factor}
\mathcal{P}_{l_\alpha\nu_\beta}(t)=
\int\! d^3\!\bp\,|\psi(\bp)|^2\!
\int\! d^3\bk\, 
\mathcal{P}_{\nu_\alpha\nu_\beta}(t)
|F_{\alpha}|^2
\,,
}
where $F_{\alpha}=(F_{\alpha j})_{m_j\rightarrow 0}$.
Notice the usual oscillation probability,
\eq{
\mathcal{P}_{\nu_\alpha\nu_\beta}(t)=
\Big|\sum_j
U_{\alpha j}e^{-iE_{\nu_j}(\bk)t}U^{\dag}_{j\beta}\Big|^2
\,,
}
factors out from the creation probability of $l_\alpha\bar{\nu}$, $|F_{\alpha}|^2$,
for massless neutrinos. Such factorization is what allows the definition of the state
Eq.\,\eqref{alphabeta} as a flavor state, since
\eq{
\label{Plnu:m=0}
\mathcal{P}_{l_\alpha\nu_\beta}(t)
\approx \delta_{\alpha\beta}\frac{\Gamma_\alpha}{\Gamma}
\,,
}
for $1/\Gamma\ll t\ll L_{\rm osc}$, where $L_{\rm osc}$ is the
typical flavor oscillation length (period). Therefore, the antineutrino flavor state
$U_{\alpha j}\ket{\bar{\nu}_j}$ is only created jointly with the charged lepton
$l_\alpha$\,\cite{flavorLee,giunti:torino04}. Notice Eq.\,\eqref{Plnu:m=0} correctly
coincides with the branching ratio of the decay $\pi\rightarrow l_\alpha\bar{\nu}$.
Neutrinos, however, are not strictly massless and we may have initial flavor
violation because different neutrino masses contribute differently to each channel
$\pi\rightarrow l_i+\bar{\nu}_j$\,\cite{flavorLee}. We will focus on initial flavor
violation and denote the interval of time satisfying $1/\Gamma\ll t\ll L_{\rm osc}$
by $t=0$.

We can make the flavor violating contributions explicit by rewriting the term inside
the square modulus in Eq.\,\eqref{Plnu} as
\eq{
\sum_{j=1}^{3}U_{\alpha j}U^*_{\beta j}F_{\alpha j}=
\delta_{\alpha\beta}F_{\alpha 1}+\sum_{j=2}^{3}U_{\alpha j}U^*_{\beta j}\Delta
F_{\alpha j}
\,,
}
where $\Delta F_{\alpha j}\equiv F_{\alpha j}-F_{\alpha 1}$.
Thus the square modulus becomes
\eq{
\label{sum23}
|\sum_{j=1}^{3}U_{\alpha j}U^*_{\beta j}F_{\alpha j}|^2=
\delta_{\alpha\beta}|F_{\alpha 1}|^2+
\delta_{\alpha\beta}2\mathrm{Re}
\Big[F_{\alpha 1}^*\sum_{j=2}^{3}U_{\alpha j}U^*_{\beta j}\Delta F_{\alpha j}
\Big]
+
\Big|\sum_{j=2}^{3}U_{\alpha j}U^*_{\beta j}\Delta F_{\alpha j}\Big|^2
\,.
}
We recognize that only the last term of Eq.\,\eqref{sum23} is flavor
non-diagonal. The second term, which is flavor diagonal, is estimated in appendix
\ref{ap:flavorcons} and shown to be much smaller than the flavor violating
contribution.

Specializing to $\alpha\neq\beta$, under the approximation of $U_{\alpha3}U_{\beta
3}^*\approx 0$ (which is valid if $\alpha=e$ or $\beta=e$), the initial creation
probability yields
\eq{
\label{Pmunue}
\mathcal{P}_{l_\alpha\nu_\beta}(0)=
\int\! d^3\!\bp\,|\psi(\bp)|^2\!
\int\! d^3\bk\,
|U_{\alpha 2}U_{\beta 2}^*|^2
|\Delta F_{\alpha 2}|^2
\,.
}
For the two family parametrization, we have $|U_{\alpha 2}U_{\beta
2}^*|^2=\frac{1}{4}\sin^2\!2\theta$, thus indicating that this phenomenon is
indeed mixing dependent.

To analyze the most dominant contribution to Eq.\,\eqref{Pmunue}, we recall that a
general function $g(x)$ can be expanded
\eq{
g(x+a)-g(x-a)\approx g'(x)2a
\,,
}
for small enough $a$.
Moreover, if $g(x)=\prod_{i=1}^n g_i(x)$, the relative difference can be written
\eq{
\frac{g(x+a)-g(x-a)}{g(x)}\approx 2a\sum_i^n \frac{g'_i(x)}{g_i(x)}
\,.
}
Taking $x$ to be $m^2_2=\overline{m^2}+\frac{1}{2}\Delta m^2$ and
$a=\frac{1}{2}\Delta m^2=\frac{1}{2}(m^2_2-m^2_1)$ we can estimate the different
contributions that compose $F_{\alpha 2}$:
\begin{enumerate}
\renewcommand{\labelenumi}{($\delta$\arabic{enumi})}
\item $g=E_{\nu_2}^{-1/2}$: \quad
$\displaystyle
a\frac{g'}{g}=-\frac{\Delta m^2}{4\bar{E}_\nu^2}$\,,
\label{delta2}
\item $g=(\Delta E_{\alpha 2}-i\Gamma/2\gamma)^{-1}$: \quad
$\displaystyle
a\frac{g'}{g}= i\frac{\Delta m^2}{2\bar{E}_\nu\Gamma}\gamma$\,,
\item $g=|\bk|_2^{\rm EC}$: \quad
$\displaystyle
a\frac{g'}{g}\sim \frac{\Delta m^2}{2\bk^2}$\,,
\item $\displaystyle\frac{|\tilde{\mathscr{M}}_{\alpha 2}^{\rm EC}|^2
-|\tilde{\mathscr{M}}_{\alpha 1}^{\rm EC}|^2}{|\tilde{\mathscr{M}}_{\alpha 2}^{\rm
EC}|^2_{m_2\rightarrow 0}}\approx
\frac{\Delta m^2}{2M^2_\alpha}
\Big(\frac{M^2_\pi+2M^2_\alpha}{M^2_\pi-M^2_\alpha}\Big)
$.
\end{enumerate}
We are assuming the energy conserving values ($\Delta E_{ij}\approx 0$),
which is an excellent approximation considering $\bar{E}_\nu,|\bk|$ are essentially
the same either if we compute it using $m^2_2$ or $m^2_1$. Conventionally we
will take the simple average $\overline{m^2}=\frac{1}{2}(m^2_1+m^2_2)$.
In particular, in $(\delta 3)$, $|\bk|_2^{\rm EC}$ denotes the momentum of neutrino
$\nu_2$, assuming energy conservation and $\bp\approx 0$: 
$|\bk|_2^{\rm EC}=\sqrt{E^2_{\nu_2}-m^2_2}$, where
$E_{\nu_2}=(M^2_\pi-M^2_\alpha+m^2_2)/(2M_\pi)$.
We also note that
$\Gamma\gg \Delta m^2/2\bar{E}_\nu$ is satisfied recalling 
$\Gamma=2.53\times 10^{-8}\mathrm{eV}$\,\cite{pdg} and 
$\Delta m^2/2\bar{E}_\nu\sim\frac{1}{6}\times10^{-7}\mathrm{eV}\frac{\Delta
m^2}{1\rm eV^2}$, where $\Delta m^2$ 
is either $|\Delta m^2_{12}|\sim 0.8\times 10^{-4}$ or $|\Delta m^2_{23}|\sim
2.5\times 10^{-3}$\,\cite{vissani:review}. This condition is necessary to have
coherent flavor neutrino creation\,\cite{flavorLee}. From $\Gamma\ll
\bar{E}_\nu,|\bk|^{\rm EC}$, it is also clear that among the different contributions
$(\delta n)$, the dominant contribution is given by $(\delta 2)$. 
A through analysis of the difference between the amplitudes $\mathscr{M}_{\alpha j}$,
estimated in $(\delta 4)$, is shown in appendix \ref{ap:pidecay}. Therefore we can
neglect all differences due to $\Delta m^2$ in $F_{\alpha j}$ except in the terms
$(\Delta E_{\alpha j}-i\Gamma/2\gamma)^{-1}$ and obtain
\eqarr{
|\Delta F_{\alpha 2}|^2&\approx&
\frac{|\mathscr{M}_{\alpha +}^{\rm EC}|^2}{N_{\alpha +}}
\bigg|
\frac{1}{\Delta E_{\alpha 2}-i\frac{\Gamma}{2\gamma}}
- \frac{1}{\Delta E_{\alpha 1}-i\frac{\Gamma}{2\gamma}}
\bigg|^2
\\
&\approx&
\label{DeltaF:final}
\frac{|\mathscr{M}_{\alpha +}^{\rm EC}|^2}{N_{\alpha +}}
\Big(\frac{\Delta m^2}{2E_\nu}\Big)^2
\frac{1}{\Big[(\Delta E_{\alpha +})^2 + \frac{\Gamma^2}{4\gamma^2}\Big]^2}
\,,
}
where the subscript $+$ means we assume $m^2_2=m^2_1=\overline{m^2}$, as well as in
$E_\nu=\sqrt{\bk^2+\overline{m^2}}$. It is also implicit that $|\mathscr{M}_{\alpha
+}^{\rm EC}|^2$ refers to $|\mathscr{M}_{\alpha j}^{\rm EC}|^2$ with
$m^2_j\rightarrow \overline{m^2}$ and without the mixing matrix element $|U_{\alpha
j}|^2$  [see Eq.\,\eqref{F:def}].
Notice we are already assuming $\bk^2\gg\overline{m^2}$, otherwise the term inside
parenthesis should be kept as $E_{\nu_2}(\bk)-E_{\nu_1}(\bk)$. Although the
$|\bk|\rightarrow 0$ limit of such term in Eq.\,\eqref{DeltaF:final} is well defined
and gives $\frac{1}{2}\Delta m^2/\sqrt{\overline{m^2}}\sim \Delta m=m_2-m_1$. In
practice, for realistic $|\bk|$, we could assume massless neutrinos for these terms.

The flavor violating creation probability in Eq.\,\eqref{Pmunue} can be calculated
in analogy to Eq.\,\eqref{tchi2:3}, using Eq.\,\eqref{Gammaij}, which gives
\eq{
\label{Pmunue:final}
\mathcal{P}_{l_\alpha\nu_\beta}(0)\approx
\frac{1}{2}\sin^2\!2\theta
\frac{\Gamma_\alpha}{\Gamma}\Big(\frac{\Delta m^2}{2E_\nu\Gamma}\Big)^2_{\rm EC}
\,,
}
where the two family parametrization, $|U_{\alpha 2}U_{\beta
2}^*|^2=\frac{1}{4}\sin^2\!2\theta$, was employed and $\bp\approx 0$ (pion at rest)
was considered by adjusting $\psi(\bp)$.
The following integral was also necessary,
\eq{
\label{int:[]^2}
\int_{-\infty}^{\infty}d\lambda
\frac{1}{\big[\lambda^2+\frac{\Gamma^2}{4\gamma^2}\big]^2}
=
\frac{2\pi}{\Gamma}\Big(\frac{2\gamma^2}{\Gamma^2}\Big)
\,.
}
One can recognize the term inside parenthesis in
Eqs.\,\eqref{DeltaF:final} and \eqref{int:[]^2} as the additional contribution that
appears in Eq.\,\eqref{Pmunue:final}.

Let us estimate some specific flavor violation probabilities (branching ratios):
\eq{
\label{Plnu:num}
\frac{\mathcal{P}_{\mu\nu_e}(0)}{\sin^2\!2\theta_{12}} \sim
10^{-9}
\,,~~
\frac{\mathcal{P}_{e\nu_\mu}(0)}{\sin^2\!2\theta_{12}} \sim
3\times 10^{-15}\frac{\Gamma_e}{\Gamma}
\,,~~
\frac{\mathcal{P}_{\mu\nu_\tau}(0)}{\sin^2\!2\theta_{23}} \sim
10^{-6}
\,.
}
To compute the last value in Eq.\,\eqref{Plnu:num}, we considered $|\Delta
m^2_{13}|\approx |\Delta m^2_{23}|\gg |\Delta m^2_{12}|$.

%%%%%%%%%%%%%%%%%%%%%%%%%%%%%%%%%%%%%%%%%%%
\section{Discussions}
\label{sec:discussion}

The important point of this detailed calculation is that lepton flavor violation
should necessarily occur when neutrinos are created because it is unlikely that the
expression in Eq.\,\eqref{Pmunue} would cancel exactly.
It is also important to emphasize that neutrinos should be detected as flavor states,
as defined (approximately) in Eq.\,\eqref{alphabeta}, to observe the flavor violation
effects. The coherent creation of neutrino flavor states is indeed guaranteed from
the observations of neutrino oscillations.
When neutrinos are not explicitly detected, their effects can be computed
from an incoherent sum of the contributions of each neutrino mass
eigenstate\,\cite{shrock}, as in the intended direct measurements of absolute
neutrino mass. Extensive investigations in such context, were first reported in
Ref.\,\onlinecite{shrock}. On the other hand, if mass eigenstates were created and
detected incoherently, flavor violating effects would be analogous to flavor changing
processes for quarks, at tree level, without the explicit appearance of the $\Delta
m^2$ dependence.

The neutrino flavor violation effects reported here are, in general, very small but
relatively larger than what would be expected from a naive estimate $\Delta
m^2/E_\nu^2$ (a similar result is indeed obtained in
Refs.\,\onlinecite{blasone:short} and \onlinecite{liliu}) because of the presence of
the finite decay width $\Gamma$, which is very small for pions. We could define,
differently from Eq.\,\eqref{alphabeta}, that the neutrinos created jointly with the
charged lepton $\alpha$ is $\nu_\alpha$ by definition\,\cite{giunti:torino04}.
However, the difference between such definition [Eq.\,(3.16) of
Ref.\,\onlinecite{giunti:torino04}a] and the usual definition in
Eq.\,\eqref{alphabeta} carries the factor $\Delta m^2/E_\nu^2$ and it is negligible
compared to the factor we have calculated in Eq.\,\eqref{Pmunue:final}. Thus the
effect calculated in Eq.\,\eqref{Pmunue:final} is dominant, even if we distinguish
the neutrinos created from different sources\,\cite{giunti:torino04}. In fact,
intrinsic neutrino flavor violation effects can not be large because otherwise there
would be no coherent creation of neutrino flavor states and there would be no flavor
oscillation\,\cite{flavorLee,kayser:81,giunti:torino04}. Of course, this analysis is
modified if there are genuine non-standard interactions\,\cite{grossman:NI}.

Previous calculations of intrinsic neutrino flavor
violation\,\cite{blasone:short,liliu,giunti:torino04} did not explicitly considered
the contribution of the finite decay width of the parent particle and either
neglected the effect\,\cite{giunti:torino04} or considered it
unphysical\,\cite{blasone:short,liliu}. The arguments of
Ref.\,\onlinecite{blasone:short} is based on a formalism that uses a unitarily
inequivalent vacuum that guarantees initial neutrino flavor
conservation\,\cite{blasone:short} but also implies slightly different oscillation
formulas\,\cite{BV}. Instead, the intrinsic flavor violation effect calculated in
Eq.\,\eqref{Pmunue:final} should be regarded as a genuine physical consequence of
massive neutrinos with mixing and it contradicts neither the weak Hamiltonian as
stated in Ref.\,\onlinecite{liliu} nor any experimental observations. The qualitative
occurrence of intrinsic neutrino flavor violation, that in the context of flavor
oscillations could be called initial flavor violation, could be anticipated in more
phenomenological calculations of flavor oscillation probabilities considering
scalar\,\cite{ccn:WP} or fermionic\,\cite{bernardini,ccn:no12} wave packets but its
magnitude could not be determined without the full consideration of the interaction
responsible for neutrino creation.

The expression in Eq.\,\eqref{sum23} reminds the $\Delta S=2$ contribution from
box diagrams in $K^0$--$\bar{K}^0$ mixing (see, \textit{e.g.},
Ref.\,\onlinecite{DGH}, p.\,235). Such contribution is suppressed by the GIM
mechanism\,\cite{gim} because it involves the sum of the contributions of quarks
$u,c$ and $t$ in the loop.
Equation \,\eqref{sum23}, however, is not loop suppressed
and, differently for quarks, the mixing angles are large. These facts explain the
relatively large effect calculated in Eq.\,\eqref{Pmunue:final}, despite tiny
neutrino mass differences. In fact, the effect is much larger than loop suppressed
effects such as the lepton flavor violating decay $\mu\rightarrow e\gamma$ in the SM.
Although, in models beyond the SM such as the MSSM, such effects can be larger than
the current experimental limit\,\cite{casasibarra}.

Despite the arbitrariness of the cutoff function $f$, the expression in
Eq.\,\eqref{Pmunue} is finite independently of the presence of that function.
This feature shows the robustness of the calculation as the cutoff scale $\Lambda$
may be chosen from a wide range without affecting the results. A expression very
similar to Eq.\,\eqref{Pmunue:final} was estimated in an unrealistic exactly solvable
QFT model of Lee-type in Ref.\,\onlinecite{flavorLee}, also showing that the
intrinsic neutrino flavor violation effects calculated here bear some universality
independently of the particular interaction in question.

The further inclusion of radiative corrections to the formalism developed in
Sec.\,\ref{sec:intro} does not seem to be straightforward. The corrections have to
be included without spoiling the conservation of probability of
Eq.\,\eqref{psi+chi}. It is also possible that deviations from the exponential
decay law would emerge from such corrections or from an approximation scheme distinct
from the WW approximation. Deviations from exponential behavior are indeed expected
for very short or very long times from the unitary evolution of quantum
mechanics\,\cite{khalfin}.
A brief connection with perturbative QFT is also shown in appendix \ref{ap:qft}.
The inclusion of finite widths in perturbative QFT is interesting
in its own right because it mixes up different orders in perturbation theory and
special care is necessary in gauge theories to keep track of gauge
invariance\,\cite{denner}. Obviously, to fully specify the dominant cutoff scale
$\Lambda$, radiative corrections should be explicitly considered. The study
of the renormalization procedure also needs careful analysis.
In this respect, it should be emphasized that the necessity of the cutoff function
$f$ is not related to the nonrenormalizability of the Fermi interaction. The
same asymptotic behavior ($|\mathscr{M}_{ij}|\sim \bk^2$) would require a cutoff
function if instead we adopted a Yukawa-type interaction which is renormalizable.
In the context of neutrino propagation and oscillation, the inclusion of finite
widths was also considered in Refs.\,\cite{grimus,beuthe} at lowest order.

Another possible application of the formalism developed in Sec.\,\ref{sec:WW}
concerns the study of the effects of the finite width to the effective size of the
decaying particles. The roles played by the finite width and the intrinsic momentum
uncertainty, encoded here in the wave function $\psi(\bp)$, are not clear but they
are crucial to the occurrence of neutrino oscillations, a phenomenon that requires
quantum coherence. As it is well known, a small uncertainty in the spatial
localization of the neutrinos are necessary to the observation of neutrino
oscillations\,\cite{kayser:81}. With such formalism, the quantum entaglement can be
also studied, differently of the static Lee-type model\,\cite{flavorLee}.
The extension to three-body decays should be also pursued since most of the decays
with neutrino creation, such as the beta decay or $\mu\rightarrow
e\bar{\nu}_e\nu_\mu$, have three decay particles. In that respect, it is
important to notice that the kinematics of a three-body decay is very
different from a two-body decay that emits monoenergetic particles when the
parent particle is at rest.

To summarize, intrinsic neutrino flavor violation should occur when neutrino flavor
states are created. The effect is the consequence of the slightly different creation
amplitudes, functions of different neutrino masses, that have to be
summed coherently. The smallness of the effect explains why neutrino flavor is an
approximately well defined concept in the SM and it is directly related to the
smallness of the neutrino mass differences. At the same time, small mass
splittings allow the coherent creation of neutrino flavor states that is required for
the phenomenon of neutrino flavor oscillations. The observation of the latter enabled
the recent progress in understanding some of the fundamental properties of neutrinos.

%%%%%%%%%%%%%%%%%%%%%%%%%%%%%%%%%%%%%%%%%%%
\appendix
%%%%%%%%%%%%%%%%%%%%%%%%%%%%%%%%%%%%%%%%%%%
\section{The cutoff function $f$}
\label{ap:cutoff}

We will show here that the contribution to Eq.\,\eqref{tchi2} coming from the second
term of Eq.\,\eqref{M+dM} is negligible for a cutoff function $f$ that obeys the
properties (P1) and (P2). We will adopt the particular function
 \eq{
|f(\bp,\bq,\bk)|^2=\frac{\Lambda^2}{(\Delta E_{ij})^2+\Lambda^2}
\,.
}
Close to the energy conserving values the contribution
of $|\delta \mathscr{M}_{ij}|^2$ is negligible compared to $|\mathscr{M}_{ij}^{\rm
EC}|^2$ as we can see in Eq.\,\eqref{deltaM/M}. To analyze the contribution of
$|\delta \mathscr{M}_{ij}|^2$ for $|\bk|\gg \Gamma$, we rewrite
Eq.\,\eqref{deltaM} as 
\eq{
\label{deltaM:2}
|\delta \mathscr{M}_{ij}|^2=
(\Delta E_{ij})^2 A_2 + \Delta E_{ij}A_1
\,,
}
and note that the coefficients $A_2,A_1$ are bounded functions of $|\bk|$. More
specifically, $E_{l_i}(\bp-\bk)-E_{\nu_j}(\bk)\approx E_{l_i}(\bp-\bk)-|\bk|$ is a
monotonically decreasing function bounded by $E_{l_i}(\bp)$ and $-|\bp|\cos\theta$.
We also notice that the term inside parenthesis in Eq.\,\eqref{tchi2:2} is bounded
as well as $|1-e^{-i(\Delta E_{ij}-i\Gamma/2\gamma)t}|^2$.
Thus, in analogy to Eq.\,\eqref{tchi2}, if we use $|f\,\delta \mathscr{M}_{ij}|^2$
instead of $|\mathscr{M}_{ij}^{\rm EC}|^2$, inside the integral in $dk$, we
recognize we have to compare
\eq{
\label{num:1}
\int_{E_{l_i}(\bp)-E_\pi}^{\infty}d\lambda\,\frac{|f
\delta\tilde{\mathscr{M}}_{ij}|^2}
{\lambda^2+\frac{\Gamma^2}{4\gamma^2}}
}
with
\eq{
\label{deno:1}
\int_{E_{l_i}(\bp)-E_\pi}^{\infty}d\lambda\,\frac{|\tilde{\mathscr{M}}_{ij}^{\rm
EC}|^2}
{\lambda^2+\frac{\Gamma^2}{4\gamma^2}}
\,,
}
where $\lambda=-\Delta E_{ij}$ and we are neglecting the neutrino masses. Taking
only the contribution of $A_2$ in Eq.\,\eqref{deltaM:2}, the ratio between
Eqs.\,\eqref{num:1} and \eqref{deno:1} is
\eqarr{
R&\lesssim&
\frac{|A_2|_{\rm max}\Lambda^2}{|\tilde{\mathscr{M}}_{ij}^{\rm EC}|^2}
\Big(\frac{2\pi\gamma}{\Gamma}\Big)^{-1}
\int_{-\infty}^{\infty}\frac{\lambda^2}
{[\lambda^2+\frac{\Gamma^2}{4\gamma^2}]
[\lambda^2+\Lambda^2]}
\\&\sim&
\frac{\Lambda\Gamma}{2(M^2_\pi-M^2_i)}
\ll 1\,,
}
assuming $\bp\approx 0$ and (P2) is valid. The contribution coming from $A_1$ is much
smaller.

%%%%%%%%%%%%%%%%%%%%%%%%%%%%%%%%%%%%%%%%%%%
\section{Calculation of Eq.\,$\text{\eqref{psi+chi}}$}
\label{ap:pi+lnu=1}

The second term of Eq.\,\eqref{psi+chi} can be rewritten as
\eq{
\int \!d^3\q d^3\rk|\chi_{ij}(\bq,\bk;t)|^2=
\label{chi2:1}
\int\!d^3\rp\,|\psi(\bp)|^2
\int\!d^3\rk \,
|\tilde{\chi}_{ij}(\bp,\bp-\bk,\bk;t)|^2
~,
}
where we used the change of variable $\bq\rightarrow \bp=\bq+\bk$ and the sum over
spins is implicit.

We can calculate, using (P1), the second integral of Eq.\,\eqref{chi2:1} 
assuming that the contribution of the second piece of Eq.\,\eqref{M+dM} is negligible
and noticing that the squared amplitude, summed over spins, after imposing
energy conservation, is a function only of the masses:
\eqarr{
\label{tchi2}
\int\!d^3\rk \,|\tilde{\chi}_{ij}|^2
&=&
\frac{|\mathscr{M}_{ij}^{\rm EC}|^2}{2E_\pi(2\pi)^3}\mathrm{Re}\!
\int\! d\Omega_k\int_0^{\infty}\!\! dk\,
\Big(\frac{k^2}{2E_{l_i}2E_{\nu_j}}\Big)
\frac{[1+e^{-\Gamma t/\gamma}-2e^{-i(\Delta E_{ij}-i\frac{\Gamma}{2\gamma})t}]}
{\displaystyle(\Delta E_{ij})^2+\frac{\Gamma^2}{4\gamma^2}}
\\&\approx&
\label{tchi2:2}
\frac{|\mathscr{M}_{ij}^{\rm EC}|^2}{2E_\pi(2\pi)^3}
\mathrm{Re}\!
\int\! d\Omega_k
\Big(\frac{k^2}{2E_{l_i}2E_{\nu_j}}\frac{dk}{d\lambda}
\Big)_{\rm EC}
\int_{-\infty}^{\infty}\! d\lambda\,
\frac{[1+e^{-\Gamma t/\gamma}-2e^{i(\lambda-E_\pi+i\frac{\Gamma}{2\gamma})t}]}
{\displaystyle(\lambda-E_\pi)^2+\frac{\Gamma^2}{4\gamma^2}}
~~~~~
\\&=&
\label{tchi2:3}
\frac{|\mathscr{M}_{ij}^{\rm EC}|^2}{M_\pi(2\pi)^3}
\mathrm{Re}\!
\int\! d\Omega_k
\Big(\frac{k^2}{2E_{l_i}2E_{\nu_j}}\frac{dk}{d\lambda}
\Big)_{\rm EC}
\frac{\pi}{\Gamma}\,
\big[1-e^{-\Gamma t}\big]
\,,
}
recalling that $\gamma=E_\pi/M_\pi$. In Eq.\,\eqref{tchi2:2}, the change of
variables $|\bk|\rightarrow\lambda=E_{\nu_j}+E_{l_i}$ was used and the lower end of
the integral was extended to $-\infty$, considering $M_\pi-M_i-m_j\gg \Gamma$.

Comparing Eq.\,\eqref{tchi2:3} with Eqs.\,\eqref{ReK} and \eqref{Gammaij},
after using $\Gamma=\sum_{ij}\Gamma_{ij}$, we see Eq.\,\eqref{psi+chi} is satisfied.

%%%%%%%%%%%%%%%%%%%%%%%%%%%%%%%%%%%%%%%%%%%
\section{Pion decay}
\label{ap:pidecay}

The effective Fermi interaction lagrangian is 
\eq{
\label{LF}
\lag_F= -2\sqrt{2}G_F
\Big(\bar{l}_i(x)\gamma^\mu L U_{ij}\nu_j(x)\Big)
J_\mu(x)
+h.c.,
}
where $L=\frac{1}{2}(1-\gamma_5)$, $\{U_{ij}\}$ denotes the MNS matrix while
$J_\mu$ is the hadronic current that in the case of pion decay reads
\eq{
J_\mu=V_{ud} \bar{u}_L\gamma_\mu d_L
\,.
}

Using $\lag_F$ we can calculate $\mathscr{M}_{ij}$ in Eq.\,\eqref{<V>}:
\eq{
\label{pi->}
\mathscr{M}_{ij}=iCU_{ij}\bar{u}_i(\bq)\sla{p}Lv_j(\bk)
\equiv iCU_{ij}\tilde{\mathscr{M}}_{ij}(\bp,\bq,\bk)
~,
}
where $C\equiv 2F_\pi G_F V_{ud}$.
We have used the chiral current relation\,\cite{DGH}
\eq{
\bra{0}\bar{u}(x)\gamma_\mu\gamma_5d(x)\ket{\pi^-(\bp)}=
-i\sqrt{2}F_{\pi}\frac{p_\mu}{\sqrt{2E_{\pi}(\bp)}}
\frac{e^{-ip.x}}{(2\pi)^{3/2}}~,
}
where $F_\pi\approx 92\rm MeV$ is the pion decay constant.
It is important to keep in mind that Eq.\,\eqref{pi->} should be calculated without
assuming energy conservation. In that case, the squared amplitude is
\eqarr{
\label{Mtilde}
\sum_{\rm spins}|\tilde{\mathscr{M}}_{ij}(\bp,\bq,\bk)|=
4(\pe{p}q_i)(\pe{p}k_j)-2(\pe{q_i}k_j)p^2~,
}
where $p^\mu\equiv (E_\pi(\bp),\bp)$, $q_i^\mu\equiv (E_{l_i}(\bq),\bq)$ and
$k_j^\mu\equiv (E_{\nu_j}(\bk),\bk)$.
If we consider energy conservation, we get the usual
\eq{
\sum_{\rm spins}|\tilde{\mathscr{M}}_{ij}(\bp,\bq,\bk)|_{\rm EC}=
|\tilde{\mathscr{M}}_{ij}^{\rm EC}|=
M^2_i(M^2_\pi-M^2_i)
+m^2_j(M^2_\pi+2M^2_i-m^2_j)
\,,
}
without neglecting the neutrino masses.
The remaining part of Eq.\,\eqref{Mtilde} that does not conserve energy can be
calculated by using $p=q+k+\delta p$, in four-vector notation, where $\delta p\equiv
(\Delta E_{ij},\bs{0})$:
\eq{
\label{deltaM}
|\delta \mathscr{M}_{ij}|^2=
(\Delta E_{ij})^2[\bp^2-(E_{l_i}-E_{\nu_j})^2]
-(M^2_i-m^2_j)2\Delta E_{ij}(E_{l_i}-E_{\nu_j})
\,.
}
It is important to estimate
\eq{
\label{deltaM/M}
\frac{|\delta \mathscr{M}_{ij}|^2}
{|\mathscr{M}_{ij}^{\rm EC}|^2}
\approx
-\frac{M^2_i}{M^2_\pi-M^2_i}\Big(\frac{2\Gamma}{M_\pi}+\frac{\Gamma^2}{M^2_\pi}\Big)
\,,
}
considering $\Delta E_{ij}\sim \Gamma$, $\bp\approx 0$, and the energy conserving
values for the rest of the terms. Numerically Eq.\,\eqref{deltaM/M} is dominated by
$\Gamma/M_\pi\sim 10^{-16}$ which is negligible and it supports why we neglected the
contribution of the terms $|\delta \mathscr{M}_{ij}|^2$ when computing the flavor
violation probability in Eq.\,\eqref{Pmunue:final}.

We can also calculate
\eqarr{
\label{Mjj'}
\sum_{\rm
spins}\tilde{\mathscr{M}}_{ij}(\bp,\bq,\bk)
\tilde{\mathscr{M}}_{ij'}^*(\bp,\bq,\bk)=
4(\pe{q_i}p)(\pe{p}\aver{k}_{jj'})-2p^2(\pe{q_i}\aver{k}_{jj'})~,
}
where $\aver{k}_{jj'}$ is given by Eq.\,\eqref{<k>}. To calculate
Eq.\,\eqref{Mjj'}, we made use of the completeness relation
in Eq.\,\eqref{sum:vvbar}. Furthermore, the mixed squared amplitude in
Eq.\,\eqref{Mjj'} can be decomposed, in analogy to Eq.\,\eqref{M+dM}, 
as
\eq{
\sum_{\rm spins}
\tilde{\mathscr{M}}_{ij}(\bp,\bq,\bk)\tilde{\mathscr{M}}_{ij'}^*(\bp,\bq,\bk)=
|\tilde{\mathscr{M}}_{i,jj'}^{\rm EC}|^2+|\delta\tilde{\mathscr{M}}_{i,jj'}|^2
\,.
}
Energy conservation (EC) assumes the neutrino four-momentum is $\aver{k}_{jj'}$.
Since the mass associated to $\aver{k}_{12}$, for example, is
$\sqrt{\aver{k}_{12}^2}=\sqrt{m_1m_2}$, i.e., the geometrical average, we can also
show that
\eq{
\label{1<12<2}
|\tilde{\mathscr{M}}_{i1}^{\rm EC}|^2 <
|\tilde{\mathscr{M}}_{i,12}^{\rm EC}|^2 <
|\tilde{\mathscr{M}}_{i2}^{\rm EC}|^2
\,,
}
for $m_1<m_2$. Equation \eqref{1<12<2} confirms that the contribution due to neutrino
mass differences in the amplitudes $\mathscr{M}_{ij}$ can be indeed neglected in
comparison to the contribution containing $\Gamma$, i.e., $(\delta2)$, when computing
Eq.\,\eqref{Pmunue:final}.

%%%%%%%%%%%%%%%%%%%%%%%%%%%%%%%%%%%%%%%%%%%
\section{Completeness relations for spinors with different masses}
\label{ap:vvbar}

To compute Eq.\,\eqref{Plnu} exactly, it is necessary to calculate mixed squared
amplitudes such as $\sum_{\rm spin}\mathscr{M}_{i1}\mathscr{M}^*_{i2}$, where the
subscripts $1$ and $2$ denote spinors involving different masses, $m_1$ and $m_{2}$.
We are interested, however, in calculating the sum over spins using a common basis
for the spin directions for the spinors $v_{\nu_1}(\bk)$ and
$v_{\nu_2}(\bk)$. (Depending on the parametrization adopted
$\bar{v}_{\nu_1}(\bk,r)v_{\nu_2}(\bk,s)\neq\delta_{rs}$)
The only basis where
that is possible is their common helicity basis. In that basis we have, with helicity
$h$,
\eq{
\label{sum:vvbar}
\sum_{h}v_{\nu_1}(\bk,h)\bar{v}_{\nu_2}(\bk,h)
=
\aver{\sla{k}}_{21}-\gamma^0\Delta\sla{k}_{21}
\,,
}
where
\eqarr{
\label{<k>}
\aver{k}_{21}&\equiv& \sqrt{m_1m_2}\big(\cosh\bar{\xi},\hat{\bk}\sinh\bar{\xi}\big)
\,,\\
\Delta k_{21}&\equiv&
\sqrt{m_1m_2}\big(\cosh\Delta\xi,\hat{\bk}\sinh\Delta\xi\big)
\,,
}
with $\bar{\xi}=\frac{1}{2}(\xi_1+\xi_2)$ and $\Delta\xi=\xi_2-\xi_1$. The usual
hyperbolic parametrization is employed,
i.e., $k_j=m_j(\cosh\xi_i,\hat{\bk}\sinh\xi_i)$, with the additional constraint
$m_1\sinh\xi_1=m_2\sinh\xi_2=|\bk|$.
Notice that Eq.\,\eqref{sum:vvbar} reduces to the usual $\sla{k}-m$ when $m_1=m_2=m$.

To calculate Eq.\,\eqref{sum:vvbar} we made use of the parametrization
\eq{
v_{\nu_j}(\bk,h)=\frac{m_j-\sla{k}_j}{\sqrt{m_j+E_j}}v_0(\bk,h)\,,
}
where $E_j=(k_j)_0$, and the completeness relation
\eq{
\sum_{h=\pm}v_0(\bk,h)v_0^\dag(\bk,h)=\frac{1}{2}(1-\gamma^0)\,.
}

%%%%%%%%%%%%%%%%%%%%%%%%%%%%%%%%%%%%%%%%%%%
\section{Flavor conserving effects}
\label{ap:flavorcons}

Let us estimate the effect of the second term of Eq.\,\eqref{sum23} which is flavor
diagonal. Comparing to Eqs.\,\eqref{DeltaF:final} and \eqref{Pmunue:final}, it is
possible that it could be relatively large, of the order of $\Delta m^2/E_\nu\Gamma$.
However, we can calculate 
\eq{
2\mathrm{Re}F^*_{\alpha 1}\Delta F_{\alpha 2}
\approx
\frac{|\mathscr{M}_{\alpha +}^{\rm EC}|^2}{N_{\alpha +}}
\frac{\Delta m^2}{2E_\nu}\frac{2\Delta E_{\alpha +}}
{\big[(\Delta E_{\alpha +})^2+\frac{\Gamma^2}{4\gamma^2}\big]^2}
\,.
}
After integration in $\lambda=-\Delta E_{\alpha +}$, the effect is non-null only
because of the lower integration limit,
$\lambda_0=m_j+E_{l_\alpha}(\bp)-E_\pi(\bp)\approx -M_\pi+M_\alpha$, is finite.
One can see the contribution will be proportional to
\eq{
\label{flavorcons}
(\frac{2\pi\gamma}{\Gamma})^{-1}
\frac{\Delta m^2}{2E_\nu}
\int_{\lambda_0}^{\infty} d\lambda
\frac{2\lambda}{\big[\lambda^2+\frac{\Gamma^2}{4\gamma^2}\big]^2}
\approx
\frac{\Delta m^2}{2E_\nu\Gamma}
\frac{1}{2\pi}\frac{\Gamma^2}{\lambda_0^2}
\,,
}
which is, in general, much smaller than $\big(\frac{\Delta
m^2}{2E_\nu\Gamma}\big)^2$.

%%%%%%%%%%%%%%%%%%%%%%%%%%%%%%%%%%%%%%%%%%%
\section{Connection with perturbative Quantum Field Theory}
\label{ap:qft}

The treatment of unstable states in pertubative Quantum Field Theory (QFT) is of
considerable interest since the majority of particles studied at high energies,
including the ones we call elementary, are unstable and observed as resonances. 
We will briefly show the connection between the formalism developed in
Sec.\,\eqref{sec:WW} with perturbative QFT.

Although the asymptotic ``in'' and ``out'' states can not be defined for an unstable
state, we can still calculate
\eqarr{
\label{psi:qft}
\psi(\bp,t)&=&
{}_0\bra{\pi(\bp)}U(t,0)\ket{\pi_\psi}_0
\,,
\\
\label{chi:qft}
\chi_{ij}(\bq,\bk;t)&=&
{}_0\bra{l_i(\bq),\bar{\nu}_j(\bk)}U(t,0)\ket{\pi_\psi}_0
\,,
}
where 
\eq{
\ket{\pi_\psi}_0=\int d^3\bp\,\psi(\bp)\ket{\pi(\bp)}_0\,,
}
and
\eq{
U(t,t')=
T\exp\Big[
-i\int_{-t'}^{t}V(t)
\Big]
\,,
}
with $T$ being the time ordered product. Recall that the $S$ matrix is given by
$U(\infty,-\infty)$.
The functions in Eqs.\,\eqref{psi:qft} and
\eqref{chi:qft} can be identified with the functions introduced in
Sec.\,\ref{sec:WW}. In particular, they obey the initial conditions of
Eqs.\,\eqref{ic:psi} and \eqref{ic:chi}.
They also obey 
\eqarr{
\label{psi:qft:dt}
i\frac{d}{dt}\psi(\bp,t)&=&
{}_0\bra{\pi(\bp)}V(t)U(t,0)\ket{\pi_\psi}_0
\,,
\\
\label{chi:qft:dt}
i\frac{d}{dt}
\chi_{ij}(\bq,\bk;t)&=&
{}_0\bra{l_i(\bq),\bar{\nu}_j(\bk)}V(t)U(t,0)\ket{\pi_\psi}_0
\,.
}
In particular, if the completeness relation in Fock space could be truncated
by $\id=\ket{\pi}\bra{\pi}+\sum_{ij}\ket{l_i\bar{\nu}_j}\bra{l_i\bar{\nu}_j}$, we
recover Eqs.\,\eqref{dpsidt1} and \eqref{dchidt1}.

%%%%%%%%%%%%%%%%%%%%%%%%%%%%%%%%%%%%%%%%%%%
\acknowledgments
This work was supported by {\em Fundação de Amparo à Pesquisa do
Estado de São Paulo} (Fapesp). The author would like to thank
Prof.~Orlando~L.~G.~Peres for pointing out Ref.\,\cite{grossman:NI} and Prof.
C.~O.~Escobar for Ref.\,\cite{khalfin}.

%%%%%%%%%%%%%%%%%%%%%%%%%%%%%%%%%%%%%%%%%%%

%%%%%%%%%%%%%%%%%%%%%%%%%%%%%%%%%%%%%%%%%%%%%%%%%

\begin{thebibliography}{99}
\bibitem{endnote1}
Recall that neutrino flavor states $\ket{\nu_\alpha}$ are superpositions of mass
eigenstates $\ket{\nu_j}$ with coefficients $U^*_{\alpha j}$ while antineutrino
flavor states $\ket{\bar{\nu}_\alpha}$ have coefficients $U_{\alpha
j}$\,\cite{giunti:torino04}.
\end{thebibliography}
\end{document}